\documentstyle[psfig,amstex]{mn}

\newif\ifAMStwofonts

\def\etal{et al.}

\ifoldfss
  \ifCUPmtlplainloaded \else
    \NewTextAlphabet{textbfit} {cmbxti10} {}
    \NewTextAlphabet{textbfss} {cmssbx10} {}
    \NewMathAlphabet{mathbfit} {cmbxti10} {} 
    \NewMathAlphabet{mathbfss} {cmssbx10} {} 
  \fi
  \ifAMStwofonts
    \ifCUPmtlplainloaded \else
      \NewSymbolFont{upmath} {eurm10}
      \NewSymbolFont{AMSa} {msam10}
      \NewMathSymbol{\upi}     {0}{upmath}{19}
      \NewMathSymbol{\umu}     {0}{upmath}{16}
      \NewMathSymbol{\upartial}{0}{upmath}{40}
      \NewMathSymbol{\leqslant}{3}{AMSa}{36}
      \NewMathSymbol{\geqslant}{3}{AMSa}{3E}

       \let\le=\leqslant
      \let\geq=\geqslant \let\ge=\geqslant
    \fi
  \fi
\fi 

\ifnfssone
  \newmathalphabet{\mathit}
  \addtoversion{normal}{\mathit}{cmr}{m}{it}
  \addtoversion{bold}{\mathit}{cmr}{bx}{it}
  \newmathalphabet{\mathbfit} 
  \addtoversion{normal}{\mathbfit}{cmr}{bx}{it}
  \addtoversion{bold}{\mathbfit}{cmr}{bx}{it}
  \newmathalphabet{\mathbfss} 
  \addtoversion{normal}{\mathbfss}{cmss}{bx}{n}
  \addtoversion{bold}{\mathbfss}{cmss}{bx}{n}
  \ifAMStwofonts
    \ifCUPmtlplainloaded \else
      \UseAMStwoboldmath
      \makeatletter
      \new@mathgroup\upmath@group
      \define@mathgroup\mv@normal\upmath@group{eur}{m}{n}
      \define@mathgroup\mv@bold\upmath@group{eur}{b}{n}
      \edef\UPM{\hexnumber\upmath@group}
      \new@mathgroup\amsa@group
      \define@mathgroup\mv@normal\amsa@group{msa}{m}{n}
      \define@mathgroup\mv@bold\amsa@group{msa}{m}{n}
      \edef\AMSa{\hexnumber\amsa@group}
      \makeatother
      \mathchardef\upi="0\UPM19
      \mathchardef\umu="0\UPM16
      \mathchardef\upartial="0\UPM40
      \mathchardef\leqslant="3\AMSa36
      \mathchardef\geqslant="3\AMSa3E

       \let\le=\leqslant
      \let\geq=\geqslant \let\ge=\geqslant
    \fi
  \fi
\fi 

\ifnfsstwo
  \DeclareMathAlphabet{\mathbfit}{OT1}{cmr}{bx}{it}
  \SetMathAlphabet\mathbfit{bold}{OT1}{cmr}{bx}{it}
  \DeclareMathAlphabet{\mathbfss}{OT1}{cmss}{bx}{n}
  \SetMathAlphabet\mathbfss{bold}{OT1}{cmss}{bx}{n}
  \ifAMStwofonts
    \ifCUPmtlplainloaded \else
      \DeclareSymbolFont{UPM}{U}{eur}{m}{n}
      \SetSymbolFont{UPM}{bold}{U}{eur}{b}{n}
      \DeclareSymbolFont{AMSa}{U}{msa}{m}{n}
      \DeclareMathSymbol{\upi}{0}{UPM}{"19}
      \DeclareMathSymbol{\umu}{0}{UPM}{"16}
      \DeclareMathSymbol{\upartial}{0}{UPM}{"40}
      \DeclareMathSymbol{\leqslant}{3}{AMSa}{"36}
      \DeclareMathSymbol{\geqslant}{3}{AMSa}{"3E}

       \let\le=\leqslant
      \let\geq=\geqslant \let\ge=\geqslant
    \fi
  \fi
\fi 

\ifCUPmtlplainloaded \else
  \ifAMStwofonts \else 
    \def\upi{\pi}
    \def\umu{\mu}
    \def\upartial{\partial}
  \fi
\fi

\title{Luminous early--type field galaxies at $z \sim 0.4$ --
II. Star--formation history and space density} \author[J. P. Willis
{\etal}] {J.P.~Willis,$^{1}$\thanks{Present address: European Southern
Observatory, Alonso de Cordoba 3107, Vitacura, Casilla 19001, Santiago
19, Chile.} P.C.~Hewett$^{2}$,
S.J.~Warren$^{3}$ and G.F.~Lewis$^{4}$\\ $^{1}$Departamento de
Astromomia y Astrofisica, P. Universidad Catolica, Avenida Vicuna
Mackenna 4860, Casilla 306, Santiago 22, Chile\\$^{2}$Institute of Astronomy,
Madingley Road, Cambridge CB3 0HA\\ $^{3}$Blackett Laboratory,
Imperial College of Science Technology and Medicine, Prince Consort
Road, London SW7 2BZ\\ $^{4}$Anglo-Australian Observatory, P.O. Box
296, Epping, NSW 1710, Australia} \date{Accepted YYYY monthname DD.
Received YYYY monthname DD; in original form YYYY monthname DD}

\pagerange{\pageref{firstpage}--\pageref{lastpage}} \pubyear{2000}

\begin{document}

\maketitle

\label{firstpage}

\begin{abstract}
We present a combined photometric and spectroscopic analysis of the
star formation history and space density of a sample of 485 luminous,
$M_V - 5 \log h < -20.5$, field early--type galaxies at redshifts $0.3
\la z \la0.6$. The observed $b_{\rm J}ori$ colours as a function of
redshift, mean absorption line strengths and [OII] 3727 emission
properties are used to constrain the star formation history of the
galaxies.  The mean star formation history of the early--type galaxy
sample is consistent with an old ($z_f>1$), passively evolving
luminosity--weighted stellar population. Twenty--one percent of the
sample possess detectable [OII] 3727 emission consistent with a low
level ($\la 1$ M$_{\odot}$ yr$^{-1}$) of on--going star formation.
Parametric and non--parametric estimates of the space density of the
sample are derived. The integrated luminosity density at $z\simeq
0.4$, allowing only for passive luminosity evolution, is in excellent
agreement with the local, ($\langle z \rangle = 0.1$), luminosity
density of early--type galaxies.  Overall, the sample properties are
consistent with a galaxy formation scenario in which the majority of
luminous field early--type galaxies formed at redshifts $z>1$ and have
largely evolved passively since the formation epoch.
\end{abstract}


\section{Introduction}

The star--formation history and space density of luminous early--type
galaxies in field\footnote{Throughout this paper, ``field'' is taken
to indicate regions outside rich galaxy clusters.} and cluster
environments at significant cosmological look--back times place strong
constraints upon theories of galaxy formation and evolution. Within
the class of hierarchical formation models, early--type galaxies are
predicted to result from a combination of major galaxy mergers (i.e.
those involving roughly equal mass progenitors) together with the
accretion of lower mass galaxies (Baugh, Cole {\&} Frenk 1996;
Kauffmann 1996).  The merger rate and thus the production rate of
early--type galaxies is predicted to be a strong function of local
galaxy density; early-type galaxies in clusters arise from an early
cosmic epoch of rapid merging and intense star formation whereas field
early--type galaxies arise from a more protracted period of merging
and star formation extending to redshifts $z \la 1$. Thus, in
hierarchical models, field early--type galaxies are predicted to
display lower mean ages and a greater dispersion in star formation
history with respect to cluster early--type galaxies. For example,
Kauffmann (1996) reports that cluster elliptical galaxies
(i.e. those located within parent halos of mass $10^{15}$M$_{\odot}$)
display luminosity weighted stellar ages between 8 and 12.5 Gyr
compared to ``field'' elliptical galaxies (i.e. those located within parent
halos of mass $10^{13}$M$_{\odot}$) displaying stellar ages between 6
and 11 Gyr. These predictions contrast with those resulting from
models in which early--type galaxies observed in both field and
cluster environments formed from the monolithic collapse of
galaxy--mass (i.e. $10^{11}-10^{12}$M$_{\odot}$) overdensities at
early cosmic times (Larson 1974).  The stellar populations in such
galaxies are predicted to have formed in an early, largely coeval,
burst of star formation and to have evolved passively since that time.

A important link between theoretical predictions and the observed
properties of early--type galaxies is achieved via studies of the
Fundamental Plane (hereafter FP) formed by such objects. Studies of
the FP for luminous early--type galaxies in clusters now extend to
redshifts $z \sim 0.5$ (e.g. Pahre, Djorgovski {\&} de Carvalho 1998;
J{\o}rgensen {\etal} 1999).  The observed evolution of the FP
parameters versus redshift in accordance with the predictions of a
passively evolving stellar population, combined with a low intrinsic
dispersion among the galaxies, imply that the majority of early--type
galaxies in rich clusters formed at some early ($z>2$), coeval epoch.
Further evidence that luminous early--type galaxies in rich clusters
are composed of uniformly old stellar populations is derived from the
observation that these galaxies display impressively uniform colours
over the redshift interval $0 < z < 1$ (Bower {\etal} 1992; Ellis
{\etal} 1997; Stanford, Eisenhardt \& Dickinson 1998). The
exceptionally narrow range of colours displayed both internally, and
between clusters, at significant cosmological look--back times ($\sim
8\,$Gyr) strongly supports an early, coeval formation epoch.

Given the relative abundance of data available for early--type
galaxies in rich clusters at redshift $z\sim0.5$, the generation of
samples of field early--type galaxies with which to test the central
predictions of galaxy formation theories is of considerable
interest. Initial attempts to form the FP relation for distant field
early--type galaxies (van Dokkum {\etal} 2001; Treu {\etal} 2001)
indicate that field early--type galaxies exhibit an FP relation
similar in the mean to that observed in rich clusters but displaying
an increased scatter about the mean relation. These studies are
limited by small sample sizes ($\sim 20$ galaxies) and include
galaxies over an extended redshift interval, $0.1<z<0.5$, leading to a
degree of uncertainty due to the need for evolutionary corrections
required to bring the galaxies to a common epoch.

The formation of early--type field galaxies via hierarchical merging
also implies that their number density should be a strong function of
look--back time (Kauffman, Charlot \& White 1996). However, the
failure to demonstrate any deficit in the number density of luminous
field spheroidal galaxies at redshifts $z < 1$ compared to local
studies places strong constraints upon the proposed epoch of
hierarchical merging (Schade {\etal} 1999).  It is important to note
however, that linking the predictions of numerical models directly to
observations is hindered by uncertainties associated with galaxy
classification (e.g. Abraham 1999).

Differences between the star--formation history of  galaxies in field
and cluster environments are likely to be augmented by the fact that
the environment of rich clusters appears to actively suppress
on--going/recent star--formation. In a study of 10 distant ($z \sim
0.5$), rich clusters Dressler {\etal} (1999) and Poggianti {\etal} \
(1999) report that 20{\%} of galaxies possess so--called
``post--starburst'', (a+k), spectra, indicative of a recent cut--off
in star--formation activity combined with an old stellar
population. The frequency of the a+k class in a comparable field
sample is considerably lower ($\la 4 {\%}$), although the significance
of this result is low given the small size of the field sample.
However, the idea that star--formation activity in rich cluster
galaxies is either suppressed (via ram--pressure stripping of neutral
gas) or rapidly increased at early times (via galaxy--galaxy
interactions and mergers) leading to exhaustion of the available gas
supply, remains compelling.

While the advent of the Sloan Digital Sky Survey has recently allowed
the definition of very large samples of early--type galaxies
(Eisenstein {\etal} 2001, Bernardi {\etal} 2002) most existing studies
of luminous early--type field galaxies over the redshift interval $0 <
z < 1$ have been confined to small numbers of objects present in
galaxy redshift surveys designed to constrain the global properties of
the galaxy population, e.g. the Canadian Network for Observational
Cosmology field galaxy redshift survey (CNOC; Yee, Ellingson \&
Carlberg 1996), the Canada France Redshift Survey (Hammer {\etal}
1997), and the AUTOFIB galaxy redshift survey (Heyl {\etal} 1997).
Given the under representation of luminous early--type field galaxies
in current galaxy surveys we have obtained multi--colour photometry
and medium resolution optical spectroscopy for a large, well--defined
sample of distant, luminous early--type field galaxies (Willis, Hewett
{\&} Warren 2001, hereafter Paper I). The observational data is of
sufficient quality to place strong constraints on the star formation
history and space density evolution of the population.  By obtaining
data for field galaxies with redshifts and luminosities comparable to
those studied in rich cluster environments we can perform an important
test of the central predictions of hierarchical models of galaxy
formation and evolution.

In Section 2 we review the compilation of a spectroscopic sample of
581 luminous field early--type galaxies (Paper I).  The
star--formation history of the sample, employing both colour--redshift
information and spectroscopic diagnostics is investigated in Section
3.  In Section 4 we calculate the space density of the galaxies and
the paper concludes by considering the implications for models of
galaxy formation and evolution in Section 5. Unless otherwise
indicated, values of $\Omega_M = 0.3$, $\Omega_{\Lambda} = 0.7$ and
$h= {\rm H_0} / 100$ km s$^{-1}$ Mpc$^{-1}$ are adopted for the
cosmological parameters describing the evolution of a model
Friedmann--Robertson--Walker universe.

\section[]{The galaxy sample}

A full description of the galaxy catalogue employed in this paper is
given in Paper I but a summary of the main points is included here for
completeness. A total of $9599$ objects were selected in an area of
220 deg$^2$ spread over 7 United Kingdom Schmidt Telescope (UKST)
fields at high Galactic latitude (Paper I; Section 2). The effective
area of the survey is 198.0 deg$^{2}$ and the surface density of
candidate early--type galaxies at the faint magnitude limit
($i<18.95$) is $\simeq 50 \, {\rm{deg}}^{-2}$.  Magnitudes, in the
natural UKST $b_{\rm J}ori$ photometric system, calibrated using
$8\arcsec$ diameter aperture magnitudes from CCD observations, are
available for all objects and satisfy the magnitude and colour
selection criteria summarised in Table 1.

\begin{table}
\caption{Photometric selection criteria.}
\label{tab_photo_select}
\begin{tabular}{ccccc}
16.5 & $<$ & $m_i$ & $<$ & 18.95 \\
2.15 & $<$ & $ b_{\rm J} - or $ & $<$ & 3.00 \\
& & $ or - i $ & $<$ & 1.05 \\
2.95 & $<$ & $ b_{\rm J} - i $ & & \\
\end{tabular}
\end{table}

The early--type galaxy sample was constructed employing a combination
of photometric and morphological criteria. The photometric selection
criteria shown in Table \ref{tab_photo_select} were chosen to
include the predicted colours of a passively evolving early--type
galaxy over the redshift interval $0.3<z<0.55$ while reducing
contamination from main sequence K-- and M--type stars. Images on each
photographic plate were classified as stellar or non--stellar on the
basis of a classification parameter (CP) computed from the surface
brightness profile (Paper I; Section 2.1). Application of a suitable
rescaling value determined for each plate expresses the CP value as
the number of standard deviations separating each object from the
centre of the stellar locus. Averaging the individual CP values over
all plates on which a particular object is detected then permits
robust separation of the stellar locus from the distribution of
galaxian objects down to the magnitude limit $i = 18.95$. The
candidate early--type galaxy sample was constructed by selecting the
set of objects that satisfy the photometric criteria described in
Table \ref{tab_photo_select} and display CP values $> 1.5$. In order
to constrain the level of incompleteness in the early--type galaxy
sample introduced by these criteria, a sample of candidate stellar
objects satsifying the above photometric criteria and CP $<1.5$ was
also constructed.

Spectroscopy of a sub--sample of $\sim 600$ galaxies was obtained
during 1997--1998 using the Two Degree Field (2dF) facility at the
3.9--m Anglo--Australian Telescope (Paper I; Section 3). The
contamination of the sample by stars is extremely small, $\sim 2\%$,
while the spectroscopic redshift completeness rate, excluding the few
stars, is very high $\ga 95\%$. Indeed, the majority of the
incompleteness in the spectroscopic identifications is due to
instrumental or astrometric problems unrelated to the intrinsic nature
of the target objects, i.e. close to $100\%$ of the galaxies in the
photometric sample are early--type galaxies with redshifts $0.25 < z <
0.63$. In addition to observations of early--type galaxy candidates, a
sample of 33 stellar candidates were included in the 1998 September
16--17 observations in order to determine directly the level of
incompleteness in the early--type galaxy sample. All of the 33 stellar
candidates display spectra consistent with that of a K-- or M--type
star, thus constraining the level of incompleteness in the early--type
galaxy sample to $<3${\%}.

The total of 581 galaxy redshifts was compiled from a number of
observing runs, including some commissioning observations where the
data quality was not ideal. In order to ensure a high degree of
uniformity the analysis in this paper is confined to 485 galaxies
observed in a single two--night run with 2dF on 1998/09/16--17. This
sample of 485 galaxies with redshifts is in turn divided into two
sub--samples: Sample A, consisting of 371 objects, contains all
spectra with reliably determined continuum shapes, and Sample B,
consisting of all 485 objects. The necessity for the two sub--samples
arose from two factors: the contamination of some of the galaxy
spectra by light from an uncovered LED (59 objects) and the
identification of a ``tail'' in the distribution of galaxy continuum
spectral properties arising from instrumental/reduction uncertainties
(55 objects). The contamination and reduction uncertainties did not
impair the assignment of redshifts but the continuum properties of the
galaxies so affected are not reliable. The investigation of
continuum--dependent properties, such as the generation of the
composite spectrum, is confined to the 371
galaxies in Sample A.

A comparison of the effective volume sampled by the spectroscopic
early--type galaxy sample to the typical space density of rich galaxy
clusters indicates that the early--type galaxy sample described in
Paper I is overwhelming drawn from the field. A direct estimate of
the effective survey volume is presented in Section \ref{sec_vacc} and
is computed as the mean accessible volume for galaxies drawn from
spectroscopic sample B occupying the redshift interval
$0.28<z<0.6$. Comparing the effective survey volume to the observed
space density of rich galaxy clusters drawn from a similar
redshift interval indicates that the anticipated fraction of cluster
early--type galaxies within the sample is $< 5-10${\%}.

One of the principal conclusions of Paper I was that virtually all the
galaxy spectra display properties that are entirely consistent with
observations of local galaxies (Paper I; Section 4). The composite
early--type galaxy continuum is strikingly similar to a local
composite early--type galaxy spectral energy distribution (SED) drawn
from the atlas of Kinney {\etal} (1996). However, there are some
identifiable differences. Eighty--one objects in Sample A (104 objects
in Sample B) possess detectable [OII] 3727 emission (Paper I; Section
4.3) and this sub--sample of galaxies exhibiting detectable [OII] 3727
emission is henceforth referred to as ``[OII]--emitting''. The larger
sub--sample of galaxies (290 in Sample A; 381 in Sample B) with no
detectable [OII] 3727 emission are henceforth referred to as
``Normal''. Possible biases against the detection of passively
evolving early--type galaxies that have experienced a recent star
formation episode are considered in detail in Section
\ref{sec_col_effect}. However, the lack of any significant colour
differences between the Normal and [OII]--emitting early--type galaxy
samples combined with the low star formation rates implied by the
observed [OII] 3727 emission fluxes suggests that no significant bias
exists. Indeed, the [OII] 3727 emission line luminosity is function is
sufficiently well constrained to reject the existence of a population
of intermediate star forming early--type galaxies at the limit of the
photometric selection criteria.

The observed number--redshift, N($z$), distribution of all early--type
galaxies with reliable redshifts (Sample B) is consistent with the
distribution predicted by applying the survey selection criteria to a
model early--type galaxy population described by a representative
spectral evolution and luminosity function model of Pozzetti {\etal}
(1996). The extreme range of rest--frame, $V$--band, absolute
magnitudes present in the sample is $-22.5 \la M_V - 5 \log h \la
-20.5$ with an inter--quartile range of 0.55 magnitudes (Section
\ref{sec_mag_est}).

The early--type galaxy photometry presented in Paper I does not
include a correction for Galactic extinction. To facilitate a
comparison with recent results in the literature (Section 4.4)
extinction corrections from Schlegel, Finkbeiner {\&} Davis (1998)
have been made to the galaxy magnitudes. Total to selective extinction
values of $A/E(B-V)$ = 4.035, 2.649 and 1.893 have been adopted to
correct galaxy magnitudes in the $b_{\rm J}$--, $or$-- and $i$--bands
respectively.

\section[]{Star formation history}

\subsection[]{Colour--redshift evolution}
\label{sec_colour}

Determining the possible range of age and metallicity consistent with
the observed evolution of broad--band colours places important
constraints upon the star formation history of the
luminosity--weighted stellar population. Most simply, to what extent
does the colour--redshift distribution for the galaxy sample conform
to the predictions of a single, evolving stellar population model?
Additional information is also present in the distribution of galaxy
properties, specifically, what are the physical implications of the
existence of the galaxy sub--samples, Normal and [OII]--emitting,
defined in Section 2 and how homogeneous are the galaxies within each
sub--sample?

The broad--band colours of the galaxy sample are compared to a range
of model stellar populations based upon the GISSEL96 spectral
synthesis code (Leitherer {\etal} 1996). Two fiducial stellar
population models are considered; an instantaneous burst and an
exponentially decreasing burst of star formation of e--folding time
scale $\tau=1\,$Gyr. Each model is described by a Scalo (1986) Initial
Mass Function (IMF) over the mass interval 0.1 to 125 M$_{\odot}$. The
colour--redshift evolution predicted for each model as a function of
formation redshift and metallicity is investigated within a spatially
flat ($\Omega_M=0.3$, $\Omega_{\Lambda}=0.7$, $h=0.7$) and an open
($\Omega_M=0.3$, $\Omega_{\Lambda}=0.0$, $h=0.7$) model universe.  At
the median redshift of the galaxy sample ($\langle z \rangle \simeq
0.4$) the age of the universe in the spatially flat and open models is
approximately $9\,$Gyr and $7\,$Gyr old respectively.

Colour--redshift loci on the $b_{\rm J}-or$ versus $or-i$ plane are
compared to observed colour--redshift values via the two--colour
probability density function $G (b_{\rm J}-or,or-i \, | \, z)$
(Appendix \ref{app_mod_col}). The function $G$ describes the
two--colour probability density of a simulated population of galaxies
at a redshift $z$, characterised by a spectro--photometric model
computed for a given formation redshift and metallicity, combined with
the photometric uncertainties in the $b_{\rm J}ori$ passbands.  Hence,
observed $b_{\rm J}ori$ colour distributions are compared to a
simulated distribution of galaxy colours described by the formation
redshift ($z_f$) and metallicity ($\log [Z/Z_{\odot}]$) of the
spectro--photometric evolution model. Following the maximum likelihood
technique of Sandage, Tammann \& Yahil (1979), one searches for
individual values of $z_f$ and $\log [Z/Z_{\odot}]$ that maximise the
combined probability density for realising the observed galaxy
colour--redshift values.

Let $p(u,v \, | \, z)$ be the differential probability density that a
galaxy of observed redshift $z$ will display colours in the interval
$u=b_{\rm J}-or$ to $u+{\rm d}u$ and $v=or-i$ to $v+{\rm d}v$, i.e.

\begin{equation}
{
p(u,v \, | \, z) \propto \frac{G(u,v \, | \,
z)}{\int_{-\infty}^{\infty} \int_{-\infty}^{\infty} G(u^{\prime},v
^{\prime}\, | \, z) H(u^{\prime},v^{\prime}) \, {\rm d}{u^{\prime}} \,
{\rm d}{v^{\prime}}}.
}
\end{equation}

{\noindent}Where $H(u,v)$ represents a two--dimensional window
function equal to unity when the galaxy colour lies within the
colour--selection boundaries (Section 2) and equal to zero when the
colour lies outside this region, i.e.

\begin{eqnarray}
\label{eqn_col_window}
H &=& 1 \;\; \mbox{  if    } \; 2.15 \le u \le 3.00 \\\nonumber
&& \,\,\, \;\; \mbox{  and  } \; 2.95 - u \le v \le 1.05 \\ \nonumber
H &=& 0 \;\; \mbox{  otherwise.} \nonumber
\end{eqnarray}

{\noindent}Equation \ref{eqn_col_window} describes the case
$E(B-V)=0$. For the case $E(B-V)>0$, early--type galaxy $b_{\rm J}ori$
photometry and photometric selection limits are adjusted
accordingly. The likelihood--term is then constructed from the
combination of individual probability density values estimated for the
$k=1, N$ galaxies in the sample for the given combination of
parameters ($z_f$, $\log [Z/Z_{\odot}]$), i.e.

\begin{equation}
{ \ln {\mathcal L} = \sum_{k=1}^{N} \ln p \, (u_k,v_k \, | \, z_k)  }
\end{equation}

{\noindent}Maximising the likelihood--term over the defined
two--dimensional parameter space equates to determining the most
probable combination of input parameters to generate the observed
colour--redshift values. Confidence intervals on the returned
parameter values may be estimated by determining the surfaces of
likelihood defined by the expression

\begin{equation}
{ \ln {\mathcal L} = \ln {\mathcal L}_{max} - {\frac{1}{2}}
{\chi}_{\beta}^{2} (M), }
\label{eqn_max_err}
\end{equation}

{\noindent}where ${\chi}_{\beta}^{2} (M)$ is the $\beta$--point of a
$\chi^2$ distribution with $M$ degrees of freedom (Efstathiou, Ellis
\& Peterson 1988). Contours describing a $4\sigma$ likelihood
difference relative to the maximum likelihood value are
evaluated. This criterion is adopted to encompass known uncertainties
in the optical/infrared colours of evolving galaxies predicted by
current population synthesis models (Charlot, Worthey {\&} Bressan
1996). Such likelihood surfaces provide constraints only on the
relative likelihood of various parameter combinations.

\begin{figure}
\psfig{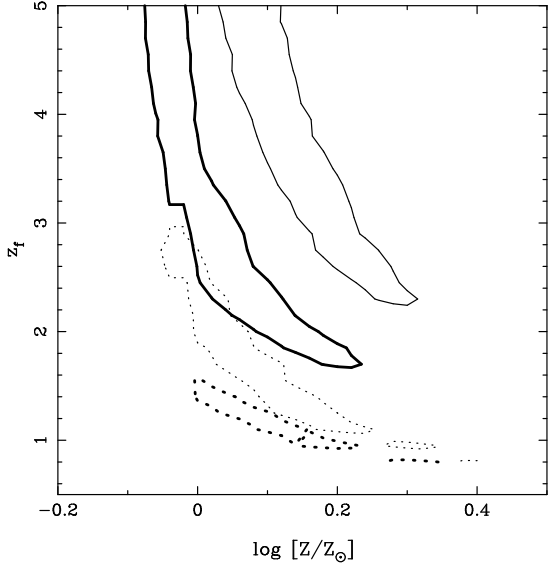}
\caption{Contours of equal likelihood in the log metallicity, $\log
[Z/Z_{\odot}]$, versus formation redshift, $z_f$, plane generated from
the colour--redshift analysis of the Normal sub--sample.  Each contour
delineates the $4\sigma$ likelihood region associated with each model;
exponentially declining burst (solid line), instantaneous burst
(dashed line), flat universe (heavy line), open universe (light line).}
\label{age_metal_red} 
\end{figure}

Figure \ref{age_metal_red} displays relative likelihood contours in
the $z_f$ versus $\log [Z/Z_{\odot}]$ plane generated for the Normal
galaxy sub--sample. The likelihood contours for each model identify a
well--defined locus of star formation scenarios ranging from young,
above--solar metallicity stellar populations ($z_f=1$, $\log
[Z/Z_{\odot}]=0.2$) to old, solar metallicity populations ($z_f>3$,
$\log [Z/Z_{\odot}]=0.0$). The properties of [OII]--emitting galaxies
are statistically identical to the Normal sub--sample; the relative
likelihood contours (not shown) are centered on the same region in the
projected $z_f$ versus $\log [Z/Z_{\odot}]$ plane but extend over a
larger area due to the smaller sample size.

Although no colour--magnitude (CM) relation for early--type galaxies
has been incorporated, the effect on the analysis is anticipated to be
negligible. In their analysis of early--type galaxies in cluster CL
1358+68 ($z\simeq0.33$), van Dokkum {\etal} (1998) report a
CM--relation $(B-V)_z = 0.866-0.018(V_z-20.7)$.  At $z=0.4$ their
redshifted $B$ and $V$ bands (i.e. subscript $z$) approximately sample
observed frame $or$ and $i$. The absolute $V$--magnitude range in our
sample is relatively small; the inter--quartile range is 0.54
magnitudes (Section \ref{sec_mag_est}). Therefore, one anticipates the
introduction of an additional dispersion term in the early--type
galaxy colours of $\sim 0.01$ magnitudes, i.e. small compared to the
colour distribution resulting from both the observed uncertainties and
estimated intrinsic dispersion (see below).

The likelihood contours presented for each model in Figure
\ref{age_metal_red} indicate that the luminosity--weighted stellar
populations describing the early--type galaxy sample are consistent
with a range of evolving stellar populations where formation redshift,
metallicity and/or star formation history vary to some degree.  Should
the galaxies form over a range of redshifts, then the
luminosity--weighted metallicity varies systematically as a function
of formation epoch, with more metal rich galaxies formed at later
epochs.  If star formation history is characterised by a sequence of
bursts then shorter bursts are necessary to match the observed
colour--redshift distribution at lower redshifts. The constraints on
the duration of bursts at high redshift are poor because the colours
of stellar populations more than $\sim 4\,$Gyr old are extremely
similar.  More complex behaviour involving variation of all three
model parameters are of course possible.

Ferreras, Charlot {\&} Silk (1999) present a markedly similar result,
from a sample of 599 morphologically classified early--type (E/S0)
cluster galaxies observed at comparable redshift and luminosity. Their
Figure 3(a) may be most directly compared to the instantaneous burst
model realised within an open universe presented in Figure
\ref{age_metal_red}. The result of Ferreras {\etal} (1999), and that
presented here, indicate that the luminosity--weighted stellar
populations of field and rich cluster galaxies were formed over a
comparable range of redshift (i.e. $z>1$) and metallicity ($\log [ Z/
Z_{\odot}]=0.0 - 0.2$).

%
%

The observed dispersion of early--type galaxy colours about a given
model locus results from the observed  photometric uncertainties and
from an intrinsic component including some combination of age,
metallicity and star formation history variations. The amplitude of
the intrinsic dispersion component for a particular colour was
estimated by first calculating the expectation value of the
distribution of galaxy colour deviations from a given model colour
locus, normalised by the photometric uncertainty, i.e. $\langle
F_{colour} \rangle$. For example, considering the $b_{\rm J}-or$
colour term one writes
\begin{multline}
F^2_{(b_{\rm J}-or), k} = {[(b_{\rm J}-or)_k - (b_{\rm
 J}-or)_{model}(z_k)]}^2 \\ / \, \sigma^2_{(b_{\rm J}-or),k},
\label{frac_expect}
\end{multline}
for a particular galaxy $k$ of observed colour $(b_{\rm J}-or)_k$,
where $(b_{\rm J}-or)_{model}(z_k)$ is the colour predicted by the
selected model locus at the redshift $z_k$ of the galaxy and
$\sigma_{(b_{\rm J}-or),k}$ is the observed photometric
uncertainty. To simplify the calculation of the colour--deviation
statistic, only galaxies that deviate from a particular colour locus
in the sense that they fall on the side of the locus opposite to the
colour--cut that defines the sample selection are included. This
limits the calculation to 270, 261 and 241 galaxies in the $b_{\rm
J}-or$, $or-i$ and $b_{\rm J}-i$ colours respectively, but ensures
that the measured dispersions are largely unaffected by the
photometric selection criteria (Table 1).

Given the normalised deviation of objects from a particular colour
locus averaged across the sample, the ``typical'' intrinsic
contribution per galaxy may be calculated by comparing the observed
deviation to the photometric uncertainty for any colour, i.e.
\begin{equation}
\sigma_{int}^2 = (F \sigma_{obs})^2 - \sigma_{obs}^2,
\label{sigma_int}
\end{equation}
where $\sigma_{int}$ is the intrinsic colour dispersion and
$\sigma_{obs}$ is the median photometric uncertainty associated with
the selected colour.

\begin{figure}
\psfig{figure=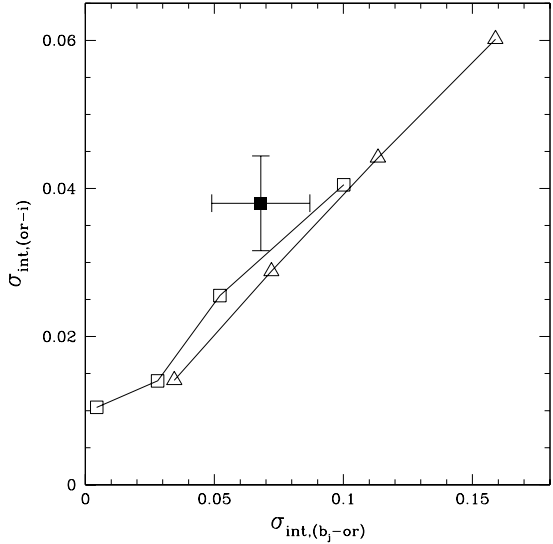,height=7.5cm,angle=0.0}
\caption{Amplitude of intrinsic variation of early--type galaxy
colours due to varying model stellar population parameters.  The solid
square with error bars indicates the intrinsic dispersion values,
calculated employing Equations \ref{frac_expect} and \ref{sigma_int},
for the galaxy sample.  The extent of the error bars represents the
variation in the intrinsic dispersion when the colour errors enclosing
68{\%} of the galaxy distribution about the median value are employed
instead of the median value.  The fiducial model consists of an
exponentially declining burst of star formation of timescale 1 Gyr
formed at redshift $z=3$ with solar metallicity.  Open triangles
indicate the colour sequence generated by varying the metallicity of
the fiducial model by $+0.05,\,0.1,\,0.15$ and $0.2$ in log
metallicity from bottom left to top right. Open squares indicate the
colour variation produced by changing the formation redshift of the
fiducial model to $z=2.7, 2.4, 2.1$ and $1.8$, again, ordered from
bottom left to top right. All simulations are considered for a
``flat'' universe model (see text). All colours are computed for model
galaxies at $z=0.4$, the median redshift of the early--type galaxy
sample.}
\label{spread_points} 
\end{figure}

Figure \ref{spread_points} compares the intrinsic spread calculated
for the $b_{\rm J}-or$ versus $or-i$ plane to colour variations
generated by a given stellar population model when varied about
central values of formation redshift and metallicity. In addition, the
dispersion resulting from varying the assumed star--formation history
between an instantaneous burst model and a model employing an
exponentially decaying burst of star formation, described by an
e--folding time scale of $\tau = 1$ Gyr, is indicated as a function of
formation redshift at fixed metallicity. Figure \ref{spread_points}
indicates that the intrinsic colour dispersion observed in the
early--type galaxy sample is consistent with only a very modest spread
in formation redshift, metallicity or star formation history. For
example, the intrinsic colour variations are consistent with an
early--type galaxy formation redshift varying over the interval
$2<z<3$ at constant metallicity.  Conversely, at fixed formation
redshift, the intrinsic colour variations are consistent with a
metallicity interval $0.0 < \log [ Z/Z_{\odot}] < 0.2$. Though more
complex scenarios involving the effect of combined variations of
formation redshift, metallicity and star formation timescale are also
allowed by the data, the low root mean square amplitude of colour
variations about a given model locus implies that the early--type
galaxy sample describes a predominantly uniform population drawn from
a narrow range of instrinsic properties.

%
%

\subsection[]{[OII] 3727 emission}
\label{sec_opt_col}

\subsubsection[]{Star formation or active galactic nuclei?}

[OII] 3727 emission is detected in 104 galaxies ($\sim 25${\%}) of
galaxies from Sample B\footnote{Although the spectra of a small number
of galaxies in sample B possess continua that may be affected by
contaminating LED emission (Section 2), this does not affect the
detectability of [OII] emission and the discussion of the [OII]
emission properties is based on galaxies in the larger Sample B.}
(Paper I; Section 4.3). Star--formation or the presence of an active
galactic nucleus could be responsible for the presence of [OII] 3727
emission.  Several lines of evidence suggest that the origin of the
emission in the majority of galaxies is star formation activity. The
majority of active galactic nuclei (AGN) exhibit relatively strong
[OIII] 5007, with the flux ratio [OIII] 5007/[OII] 3727 $\ga 3$ (e.g.
Baldwin, Phillips \& Terlevich 1981). However, a composite spectrum of
the 104 galaxies with detectable [OII] 3727 emission shows no evidence
for the presence of the [OIII] 4959 5007 doublet or of H$\beta$ in
emission. Indeed, in the entire galaxy sample there are no individual
detections of [OIII] 5007 or H$\beta$ in emission. The flux limit for
detection of [OIII] 5007 in the galaxy spectra is a factor $\simeq 1.5$
poorer than for [OII] 3727, due primarily to the higher sky--background
at $\sim 7000$\AA. The lack of detections of [OIII] 5007 is thus
surprising if the line ratios were typical of AGN but is entirely
consistent with the identification of the source of [OII] emission as
low--excitation star formation where the [OIII] 5007/[OII] 3727 ratio
is often unity or significantly smaller.

An estimate of the fraction of the galaxy sample hosting AGN can be
made from the number of galaxies that are radio sources. The excellent
astrometric properties of the galaxy sample (Paper I; Section 2) and
the FIRST radio survey (Becker, White \& Helfand 1995) allow unambiguous
identification of galaxies with radio sources using a matching radius
of only $2\,$arcsec. The FIRST survey includes a significant proportion
of the three UKST equatorial fields included in the full photometric
catalogue of $9599$ candidate galaxies. Of 3170 galaxy candidates
within the FIRST survey region, 95 lie within $2\,$arcsec of a FIRST
source. A further 6 galaxies posses multiple FIRST sources, within a
$10\,$arcsec radius, that strongly suggest the galaxy is the origin of
radio--lobe emission.  Thus, only 101 of 3170 ($\simeq 3\%$) of the
galaxies appear to be radio sources to $1\,$mJy at $1.4\,$GHz, compared
to the much larger percentage showing evidence for [OII] 3727
emission.  The FIRST survey does not cover the southern fields where
the 2dF spectroscopy was obtained but the NVSS (Condon {\etal} 1998)
extends to declination $-40^{\circ}$, including sources to $2.5\,$mJy
at $1.4\,$GHz. Unfortunately, the positional accuracy of the NVSS is
such that a $15\,$arcsec pairing radius is necessary when matching the
catalogue to the galaxy sample. Comparison of the matches between the
galaxy sample and the NVSS and FIRST surveys in the three equatorial
regions shows excellent agreement. NVSS includes all but the galaxies
corresponding to the faintest FIRST sources with the addition of an
extra $\sim 30\%$ of spurious ``matches'' due to the much larger
pairing radius employed. Applying the same matching radius to the
galaxies included in the 2dF spectroscopy produces 14 galaxy--NVSS
matches out of 421 galaxies within the NVSS survey area. Ten of the
NVSS--galaxy matches are expected to represent physical association,
with 4 galaxies paired by chance. The galaxies in the spectroscopic
sample thus show a very similar incidence of radio emission to that
inferred using the FIRST survey.  Of the 14 NVSS matches none are
included among the 104 [OII] 3727 emission line objects and the
composite spectrum of the 14 objects shows no evidence for any
anomalous features. Thus, based on the observed limits to the flux
ratio of the [OII] 3727 and [OIII] 4959 5007 emission lines and the
small incidence of radio emission among the galaxy sample we conclude
that star--formation activity, rather than AGN, is responsible for the
majority of the [OII] 3727 emission detected in the galaxy spectra.

\subsubsection{The [OII] 3727 luminosity function and evolution of the SFR}

[OII] 3727 emission arises within HII regions ionised by populations of
massive, short--lived O-- and B--type stars (Hogg {\etal} 1998;
Osterbrock 1989). Given the short lifetime of such populations ($\sim
10^6$ yrs), [OII] 3727 emission offers an almost instantaneous snapshot
of the current star--formation rate on cosmological
time--scales.\footnote{ While this statement is qualitatively true, the
calculation of the true star formation rate requires further
information, including the Initial Mass Function. The strength of [OII]
3727 emission resulting from an ionising field is also sensitive to the
physical state of the interstellar medium -- particularly the electron
density and metallicity.} The luminosity of an [OII] 3727
emission line of flux $f$ in a galaxy observed using 2dF at redshift
$z$ is
\begin{equation}
{
{\rm{L}}_{{\rm[OII]}} = 4 \pi d_L^2(z) ( f/A ), 
}
\label{eqn_l_OII}
\end{equation}
where $d_L(z)$ is the luminosity distance corresponding to a redshift
$z$ in the specified cosmological model. $A$ is the fraction of total
source flux intercepted by a 2dF fibre. The form of the correction for
the fraction of [OII] 3727 emission intercepted by a fibre is based on
the assumption that the distribution of continuum and [OII] 3727
emission in the galaxies is co--spatial. The observed emission line
fluxes were measured by fitting a Gaussian model to each
continuum--subtracted line (Paper I; Section 4.3).

Kennicutt's (1992) relation, derived from a sample of local galaxies,
between mean [OII] 3727 luminosity and star--formation rate (SFR);

\begin{equation}
{
\rm{SFR} (\rm{M}_{\odot} \rm{yr}^{-1}) = \frac{L_{[OII]}}{5 \times
10^{41}} \; ({\rm ergs} \; {\rm s}^{-1}).
}
\label{eqn_ken}
\end{equation}
is used to estimate the SFR.  [OII] 3727 emission luminosity, rather
than [OII] 3727 equivalent width, is employed because, while equivalent
width provides a more straightforward indicator of emission relative to
the ionising continuum, any sky--subtraction errors introduce
significant uncertainties into measurements of equivalent width.

\begin{figure}
\psfig{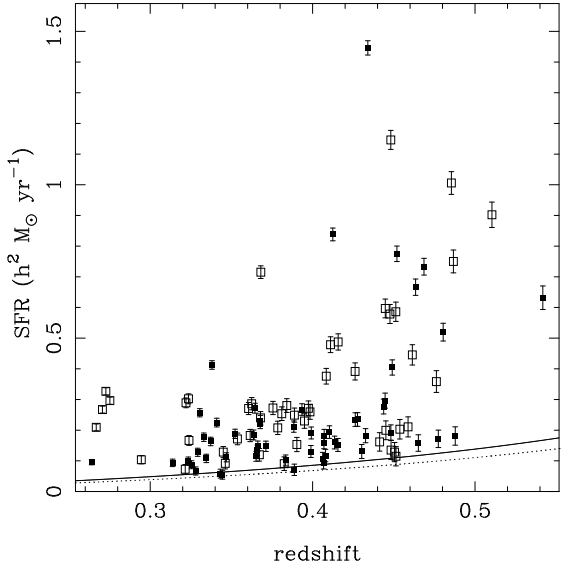}
\caption{SFR, estimated from [OII] 3727 luminosity, versus redshift for
104 galaxies with detectable emission. Open squares represent data
obtained on Night 1 (1998/9/16), filled squares represent data obtained
on Night 2 (1998/9/17). The solid and dotted lines indicate the
detection limit for nights 1 and 2 respectively.}
\label{plot_sfr} 
\end{figure}

Figure \ref{plot_sfr} displays the distribution of SFR versus
redshift, demonstrating that SFRs inferred from [OII] 3727
luminosities are small ($\la  1$ M$_{\odot}$ yr$^{-1}$). The evolution
of [OII] 3727 emission, and by implication the SFR, as a function of
redshift may be investigated by calculating the [OII] 3727
emission--line luminosity function for the galaxy sample at different
redshifts. The nature of the sample is well--suited to the application
of the $1 / \, V_{acc}$ method (Avni \& Bahcall 1980). The maximum
accessible volume associated with each [OII]--emitting galaxy in the
sample is calculated by integrating the co--moving volume element per
unit solid angle, ${\rm{d}}V / \, {\rm{d}}z$, multiplied by the
photometric selection probability as a function of redshift, $W(z)$
(Equation \ref{eqn_wz}), over the redshift interval within
which the estimated [OII] 3727 luminosity exceeds the detection
threshold for the identification of [OII] 3727 emission in the galaxy
spectrum, i.e.
\begin{equation}
{
V_{acc} = c \; {\rm{d}}\Omega
\int_{z_{min}}^{z_{({\rm{L_{[OII]}}}\geq {\rm{L_{lim}}}(z))}}
{\frac{{\rm{d}}V}{{\rm{d}}z} \, W(z) \, {\rm{d}z},}
}
\label{eqn_acc_OII}
\end{equation}
The factor d$\Omega$ scales the accessible volume by the solid angle
covered by Sample B (d$\Omega \simeq 10.2$ deg$^2$) and the factor $c$
corrects for redshift incompleteness and is equal to the inverse of
the redshift completeness rate as a function of apparent
$i$--magnitude (Paper I; Section 3.5). The cumulative space density of
[OII]--emitting galaxies displaying $\rm{L_{[OII]}} > L$,
$\Phi_{\rm{[OII]}}(>{\rm{L}})$,  is obtained by summing the inverse
accessible volume of each galaxy.

\begin{figure}
\psfig{figure=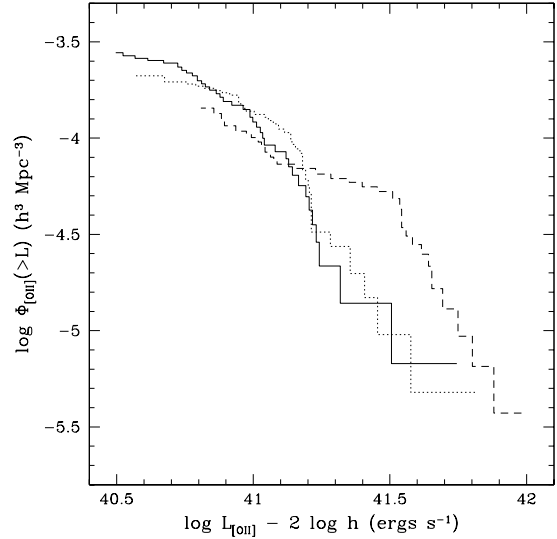,height=7.5cm,angle=0.0}
\caption{Cumulative space density of [OII]--emitting galaxies
versus [OII] 3727 luminosity. Data for three redshift shells $0.28<z<0.37$
(solid line), $0.37 \le z <0.43$ (dotted line) and $0.43 \le z < 0.6$ (dashed
line) are shown.}
\label{vacc_OII} 
\end{figure}

Evolution of the early--type galaxy SFR as a function of redshift was
investigated by computing the cumulative space density of
[OII]--emitting galaxies as a function of $\rm{L_{[OII]}}$ in three
redshift shells ($0.28<z<0.37$ (36 galaxies), $0.37 \le z < 0.43$ (32
galaxies), and $0.43 \le z < 0.6$ (31 galaxies)), chosen to include
approximately equal numbers of galaxies. The photometric selection
function was computed employing a model described by exponentially
decaying burst of star formation of e--folding time scale
$\tau=1\,$Gyr formed at $z_f=3$ with metallicity $\log
[Z/Z_{\odot}]=0.0$ in a spatially flat universe ($\Omega_M=0.3$,
$\Omega_{\Lambda}=0.7$, $h=0.7$). The enclosed volume associated with
each [OII]--emitting galaxy, $V_{enc}$ (Section 4.2), was also
calculated. The
resulting cumulative distributions, Figure \ref{vacc_OII}, show a
marked increase ($\Delta \log \Phi \sim 0.5$) in the space density of
bright ($\rm{L_{[OII]}} > 10^{41}$ ergs s$^{-1}$) [OII] 3727--emitters
at redshifts $z>0.43$.  Any difference in the luminosity function
between the two lower redshift shells depends on the exact placement
of the redshift boundary (at $z\sim0.37$) and a larger sample is
required to draw quantitative conclusions. However, the significant
increase in the space density of bright ($\rm{L_{[OII]}} > 10^{41}$
ergs s$^{-1}$) [OII]--emitters at redshifts $z>0.43$ is robust and
does not vary significantly as the $z\sim0.43$ shell boundary is
perturbed.

\subsubsection[]{The effect of star--formation on the colour
selection}
\label{sec_col_effect}

The conclusion that the fraction of early--type galaxies possessing
bright ($\rm{L_{[OII]}} >10^{41}$ ergs s$^{-1}$) [OII] 3727 emission
increases significantly for redshifts $z \ga 0.4$ is a potentially
important result (Section \ref{sec_sf_conc}). However, it is
necessary to verify that the presence of star formation activity does
not lead to a bias in the selection of galaxies as a function of
redshift that might produce a spurious evolutionary trend.

Consider a galaxy with a passively evolving stellar population formed
in an exponentially--decaying burst of star formation ($\tau=1\,$Gyr),
with solar metallicity, initiated at redshift $z_f = 3$, that later
experiences a burst of star--formation, occurring at $z_b \sim 1$. For
the adopted cosmological parameters, the age of the burst at the epoch
of observation ($z \simeq 0.4$) is $4\,$Gyr. The star--burst is assumed to
involve a population of solar metallicity with a Scalo IMF and an
e--folding time--scale $\tau = 1\,$Gyr. Model predictions for mass
fractions equal to 10, 5, 1 and 0.1{\%} of the total stellar mass in
the galaxy were generated using the GISSEL96 code.

\begin{figure}
\psfig{figure=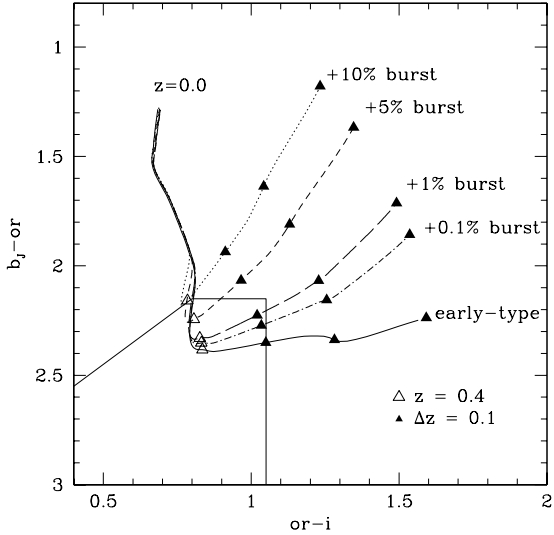,height=7.5cm,angle=0.0}
\caption{Predicted $b_{\rm J}-or$ versus $or-i$ colours as a function of
redshift for model early--type galaxies experiencing a star--burst at
redshift $z_b = 1$ (see text for details). Symbols along the tracks are
at increments $\Delta z = 0.1$. The tracks terminate, on the right, at
$z_{max}=0.6$. Note the colour ``synchronisation'' at $z \sim 0.3$ for
all models.}
\label{model_burst}
\end{figure}

Figure \ref{model_burst} shows the predicted behaviour of a galaxy
experiencing a burst of star--formation, as described above, in the
$b_{\rm J}-or$ versus $or-i$ plane as a function of redshift.
Modest star--bursts (0.1{\%} and 1{\%} of the total stellar mass)
display colours over the redshift interval $0.3 \la z \la 0.55$ that
are very similar to the underlying model early--type galaxy population
($\Delta (b_{\rm J}-or) \la 0.15$). More massive star--bursts (5{\%} and
10{\%}) produce markedly bluer colours at redshifts $z \ga 0.3$.
However, it is a feature of all rapidly fading burst models that
optical colours ``synchronise'' to those of an old stellar population
at post--burst ages $t_b \sim 4 - 5$ Gyr ($z \sim 0.25$ within the
current model), explaining the tight locus formed by the models at
$z<0.3$.

A star--burst at $z_b \sim 1$ causes the early--type galaxies to become
bluer in $b_{\rm J}-or$, moving the galaxies towards the selection
boundaries, with the effect most pronounced at the highest redshifts.
Thus, any burst of star--formation corresponding to $\ga 1\%$ of the
total stellar mass would result in early--type galaxies at redshifts
$z>0.4$ becoming lost from the sample.  We conclude that the observed
increase in the fraction of galaxies with high [OII] 3727 at redshifts
$z \ga 0.4$ is robust and can only increase in the event that a
fraction of early--type galaxies experienced massive bursts.

The similarity of the distributions of the Normal and [OII]--emitting
galaxies in the colour--redshift plane, combined with the low--level of
the [OII] 3727 emission present, constrains the fraction of the stellar
mass involved in a star--burst at $z_b \sim 1$ for the majority of the
sample to be $\la 0.1\%$. However, the degree of bluing is a strong
function of the mass of the burst and it is not possible to exclude the
possibility that some fraction of early--type galaxies experience
massive $\ga 5\%$ bursts of star formation at $z_b \sim 1$. Redshift $z
\sim 1$ is the upper limit for which useful constraints may be placed
upon star--burst activity using the current sample. Figure 4
illustrates why the signature of bursts occurring at higher redshift
are difficult to identify due to the strong degree of colour
synchronisation that takes place some $4\,$Gyr after the bursts.
Conversely, the limits on the mass of bursts of star--formation
occurring at redshifts $z \la 1$ are even tighter than the $\sim 0.1\%$
limit deduced for a burst redshift of $z_b = 1$.

\subsection[]{Absorption line indices}
\label{sec_absorb_index}

The measurement of spectral absorption line indices of old stellar
populations in early--type galaxies provides a probe of both relative
age and metallicity of the luminosity weighted stellar population.
Absorption line measures are computed using the Lick/IDS line strength
indices (Worthey {\etal} 1994). The dependence of the Lick/IDS indices,
defined by central bandpass and adjacent pseudo--continua, on the age
and metallicity of the underlying stellar population have been
investigated extensively (Worthey 1994).

Absorption line strengths in the composite spectra of the Normal and
[OII]--emitting galaxies were compared. The composite spectrum of each
sub--sample was matched to the resolution of Lick/IDS spectra via
smoothing with a Gaussian filter of wavelength--dependent FWHM (see
Appendix A of Worthey \& Ottaviani, 1997). This ensures that the
indices measured are on the same scale as those from the Lick/IDS data,
to within a small additive constant. The dependence of the indices on
galaxy velocity dispersion is also a potential concern.  Kuntschner
(2000) attempts to correct for galaxy velocity dispersion in the
measurement of Lick/IDS indices on early--type galaxies and report
corrections $\la$5{\%} at velocities $\la300\,$kms$^{-1}$ in the value of
the H$\beta$ and C4668 indices employed in this paper (see below).
Given that the velocity dispersion correction is small and is difficult
to measure accurately without stellar calibration exposures, we do not
attempt such a correction here. Errors in the measured absorption
indices measured were determined via a bootstrap procedure: absorption
indices were measured in composite spectra drawn (with replacement)
from the Normal sub--sample. A determination was made for each
sub--sample realisation, with the number of galaxies contributing to
each realisation equal to the number in each of the galaxy
sub--samples.  One--$\sigma$ errors were calculated from the resulting
distributions of absorption indices measured for 1000 such
realisations.

The models of Worthey {\etal} (1994) predict index values for a
passively--evolving, instantaneous--burst stellar population of given
age and metallicity. Kuntschner \& Davies (1998) advocate the use of
the C4668 and Balmer line indices to form a (relatively) independent
estimate of the metallicity and age respectively of early--type
galaxies in the Fornax cluster. However, they note that galaxy
properties inferred from absorption indices should be regarded solely
as {\em relative} measures.

\begin{figure}
\psfig{figure=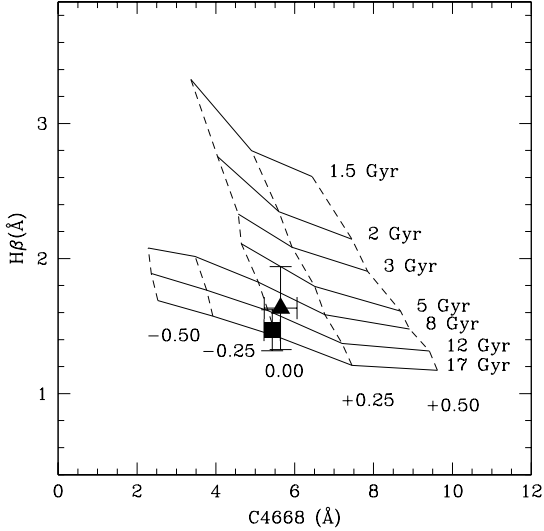,height=7.5cm,angle=0.0}
\caption{Comparison of C4668 and H$\beta$ Lick/IDS absorption indices
measured for the composite spectrum for Normal (square) and [OII]--emitting
galaxies (triangle) to indices measured from
synthetic stellar populations of varying age and
metallicity. 1$\sigma$ errors are derived via a bootstrap analysis
(see text). Stellar population predictions are derived from the models
of Worthey (1994). Labels placed at the bottom of each grid indicate
$\log [ Z / Z_{\odot} ]$, labels placed to the right of each grid
indicate stellar population age.}
\label{ew_grid1}
\end{figure}

Figure \ref{ew_grid1} shows the C4668 and H$\beta$ indices for
Normal and [OII]--emitting early--type galaxies drawn from Sample A.
The observed offset in the strength of the H$\beta$
index between the two sub--samples is in the opposite sense to that
which would be expected if the [OII]--emitting galaxies exhibited
low--luminosity nebular HII emission, effectively reducing the
strength of H$\beta$ in absorption and leading to overestimation of the
apparent age of the galaxy. One may therefore conclude that any offset
caused by nebular HII emission in the early--type galaxy sample as a
whole is smaller than the estimated error in the H$\beta$ index
determined for the [OII]--emitting galaxies (or alternatively that it
is unrelated to [OII] 3727 emission). The absorption line properties of
each sub--sample are statistically identical, indicating that the
luminosity weighted stellar populations of Normal and [OII]--emitting
galaxies share a common star formation history.

\subsection[]{Conclusions}
\label{sec_sf_conc}

To summarise, the mean colour--redshift distribution of the sample is
consistent with a luminosity--weighted stellar population that formed
at redshift $z>1$. No significant star formation has occurred since that
time. If the population formed over an extended period then there must
have been a well--defined tight relationship between the mean
metallicity of the stars and the formation epoch in order for the
galaxy population as a whole to exhibit such a small dispersion in
observed properties.  The exact relation between formation redshift and
metallicity is dependent upon the assumed star formation rate as a
function of time and the cosmological model.

The strength of the mean absorption line indices of both the Normal
and [OII]--emitting sub--samples indicates that each sub--sample is
consistent with an old ($\sim12$ Gyr), approximately solar metallicity
($\log [Z/Z_{\odot}] \simeq 0.0$ dex) luminosity--weighted stellar
population. The results of the colour--redshift and absorption line
analyses are consistent and indicate that, modulo the assumed star
formation rate as a function of time, the mean luminosity weighted
stellar population in luminous field early--type galaxies formed at a
redshift $z \gg 1$ with approximately solar metallicity.

Observation of [OII] 3727 emission in $\sim25${\%} of the galaxies
indicates that these galaxies are experiencing low SFRs of only $\la 1
\, h^2 \, {\rm{M}}_{\odot} \, {\rm{yr}}^{-1}$. This low level of star
formation is consistent with the results of the colour--redshift and
absorption index analyses that indicate that Normal and [OII]--emitting
early--type galaxies possess essentially identical luminosity--weighted
stellar populations. The space density of the brightest [OII]--emitting
galaxies displays an increase of a factor $\sim 3$ at the largest
redshifts. However, the colour selection criteria that define the
sample mean that recent, $z_b \simeq 1$, star formation events must
involve $< 0.1${\%} of the total stellar mass of the galaxies.

The conclusions regarding the star formation history of luminous field
early--type galaxies appear to be inconsistent with the hierarchical
merging of gas rich, massive galaxies at redshifts $z \la 1$. Rather,
the observations are more consistent with the accretion of gas rich,
low mass (presumably dwarf) galaxies leading to a small burst of star
formation superimposed upon a dominant, quiescent stellar population.
The increase in the space density of the brightest [OII]--emitting
galaxies at large redshift may indicate an increase as a function of
redshift in the accretion rate of low mass galaxies. However, such
a conclusion is critically dependent on the identification of the
galaxy population studied here as the progenitors of early--type
galaxies today. If in fact the galaxy population studied here 
represents only a small fraction of the present--day population of
early--type galaxies then the conclusions concerning the formation 
history outlined above could be extremely misleading.

The importance of progenitor--bias in the context of the evolution of
early--type galaxies in clusters has been stressed by van Dokkum \&
Franx (2001). In the context of the sample of field early--type
galaxies it is essential to demonstrate that the space density of the
population at $z \simeq 0.4$ is comparable to the space density of
early--type galaxies today and this question is addressed in Section
4.

\section[]{The luminosity function}

The evolution of the space density of early--type galaxies as a
function of redshift places strong constraints upon the extent to which
the early--type galaxy population formed via the hierarchical merging
of galactic ``sub--units''. The 485 galaxies of Sample B allow a direct
estimate of the luminosity function at $z\simeq 0.4$ to be made. A
parametric estimator allows a comparison with parametric descriptions
of luminosity functions from other samples while a non--parametric
estimator provides a direct determination of the luminosity function
without the imposition of a possibly inappropriate model
representation. Both types of estimator are employed here.

\subsection[]{Absolute Magnitudes}
\label{sec_mag_est}

\begin{figure}
\psfig{figure=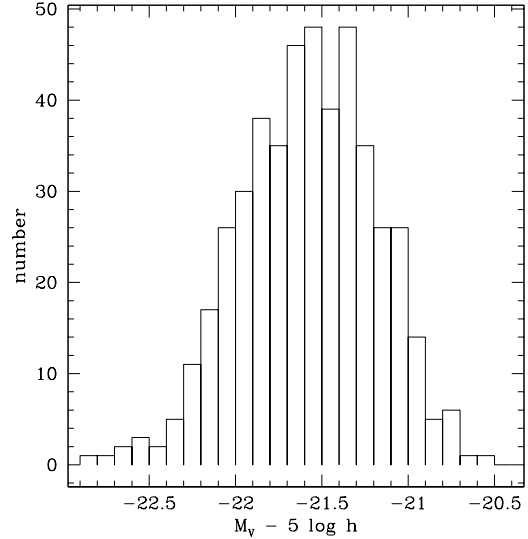,height=7.5cm,angle=0.0}
\caption{Rest--frame absolute $V$--band magnitudes, corrected for
passive luminosity evolution, at a common epoch, $z=0.4$. See text for
discussion.}
\label{abs_mag_hist}
\end{figure}

The analysis of the colour--redshift distributions for each galaxy
sub--sample in Section 3.2 demonstrated that the observed colours are
consistent with the predictions of a single passively--evolving stellar
population model, modulo a small intrinsic dispersion. The effects of
passive luminosity evolution over the redshift interval $0.3 \le z \le
0.6$ are significant and the derived rest--frame absolute magnitudes
need to be corrected for passive stellar luminosity evolution.  In
addition, referring luminosity function parameters to a common epoch
provides a necessary reference point against which complementary
studies of early--type galaxy may be compared. To correct the apparent
$b_{\rm J}ori$ magnitudes of each galaxy to a common epoch it is
necessary to compute the apparent magnitude of that galaxy as a
function of redshift, i.e.,
\begin{multline}
m_n(z) = m_n - A_n - 5 \log [d_L(z_n)/d_L(z)] \\
- ([e+k](z_n)-[e+k](z)) \, ,
\label{eqn_magz}
\end{multline}
where each galaxy with magnitude $m_n$ and redshift $z_n$ is corrected
for Galactic extinction, $A_n$, the effect of differential luminosity
distance as a function of redshift, $d_L(z)$, and a differential
evolution plus $k$--correction term as a function of redshift
$[e+k](z)$ appropriate to each photometric passband. Rest--frame
absolute $V$--band magnitudes are then determined from apparent
$i$--band magnitudes computed at a common redshift, $z_c=0.4$,  i.e.,
\begin{equation}
{ M_V = m_n(z_c) - 25 - 5 \log d_L(z_c) - k(z_c) + (V-i)_{z=0.4}, }
\label{eqn_abs}
\end{equation}
where $k(z_c)$ is determined from the SED of a $z=0.4$ galaxy generated
by the specified spectral evolution model and $(V-i)_{z=0.4}$ is the
rest frame $V-i$ colour for an early--type galaxy at $z=0.4$. Figure
\ref{abs_mag_hist} shows the absolute magnitude distribution
computed for an exponentially decaying burst of star formation of
e--folding time scale $\tau=1\,$Gyr formed at a redshift $z_f=3$ with
solar metallicity within a spatially flat
universe (see Table \ref{tab_lf_models}:  ``Model 2'').
 
\subsection[]{The luminosity function: ${1 \, / \, V_{acc}}$
estimation}
\label{sec_vacc}

\begin{figure}
\psfig{figure=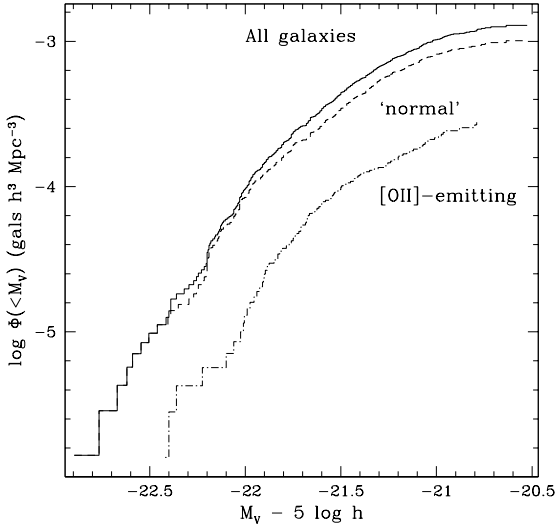,height=7.5cm,angle=0.0}
\caption{Cumulative luminosity functions corresponding to each
spectral sub--class compared to the combined galaxy sample. Each
luminosity function was generated employing ``Model 2'', i.e. an
exponentially decaying burst of star formation at $z_f = 3.0$ with an
e--folding time scale of $\tau=1\,$Gyr within a spatially flat
universe.}
\label{clf_type}
\end{figure}

In the $1 / V_{acc}$ estimator (Avni \& Bahcall 1980) the maximum
accessible volume associated with each galaxy in the sample is
computed by integrating the co--moving volume element per unit solid
angle, ${\rm{d}}V / \, {\rm{d}}z$, multiplied by the probability that
a galaxy of given magnitude and colour enters the sample as a function
of redshift, $W(z)$ (Equation \ref{eqn_wz}), over the
specified redshift limits, i.e.
\begin{equation}
{
V_{acc} = c \; {\rm{d}}\Omega \int_{z_{min}}^{z_{max}}
\frac{{\rm{d}}V}{{\rm{d}}z} \, W(z) \, {\rm{d}}z,
}
\label{eqn_acc}
\end{equation}
The factor d$\Omega$ scales the accessible volume by
the appropriate solid angle and the factor $c$ corrects for redshift
incompleteness. The enclosed volume associated with each galaxy,
$V_{enc}$, is calculated by replacing the integration limit $z_{max}$
by $z_{gal}$.  Redshift limits of $z_{min}=0.28$ and $z_{max}=0.60$
were adopted to exclude galaxies at redshifts where the probability
$W(z)$ becomes very small. The resulting number of galaxies in the
Normal and [OII]--emitting sub--samples were 367 and 99 respectively.
The cumulative $1/V_{acc}$ luminosity function, $\Phi(<M)$, is
calculated by summing the inverse accessible volumes for each galaxy
(Figure \ref{clf_type}).

\begin{table}
\caption{Spectral evolution models and cosmological parameters
employed in luminosity function analyses. For exponentially decaying
star formation histories, the e--folding time scale is set to
$\tau=1\,$Gyr.  The ``non--evolving'' spectral evolution model employs
the Kinney SED to determine the $k$--correction term. Although
$h$--dependent values are normalised to $h=1$ the adopted value of
$h=0.7$ determines the age--redshift relation.}
\label{tab_lf_models}
\begin{tabular}{clcccc}
Model & \multicolumn{3}{c}{Spectral model} &
\multicolumn{2}{c}{Cosmological model} \\
 & SF history & $z_f$ & $\log [Z/Z_{\odot}]$ & $\Omega_M$ &
 $\Omega_{\Lambda}$ \\
1 & Instantaneous & 1.4 & 0.0 & 0.3 & 0.7 \\
2 & Exponential   & 3.0 & 0.0 & 0.3 & 0.7 \\
3 & Non--evolving & --  & --  & 0.3 & 0.7 \\
4 & Instantaneous & 2.6 & 0.0 & 0.3 & 0.0 \\
5 & Exponential   & 4.0 & 0.1 & 0.3 & 0.0 \\
6 & Non--evolving & --  & --  & 0.3 & 0.0 \\
\end{tabular}
\end{table}

Internal consistency was verified by comparing cumulative luminosity
functions for the two sub--samples and for each 2dF field using a
two--sample Kolmogorov--Smirnov (KS) test. Figure \ref{clf_type} shows
cumulative luminosity functions for the Normal and [OII]--emitting
sub--samples together with the combined sample. A KS--test applied to
the Normal and [OII]--emitting sub--samples gives $D_{max}$ and
associated probabilities that the samples are drawn from the same
distribution, of: (0.11, 0.34), indicating no significant differences.
$D_{max}$ and associated probabilities for samples from each 2dF field
compared to the remaining three fields are (0.10,0.36), (0.13,0.10),
(0.06,0.90), (0.12,0.16) for fields 1, 2, 3, and 4 respectively. The
sample thus shows no evidence for significant differences in the
luminosity function from field to field, although the effects of cosmic
variance are likely responsible for the smaller probabilities
associated with fields 2 and 4.

The effective volume probed by the spectroscopic early--type
galaxy sample is computed as the mean accessible volume of all
galaxies drawn from Sample B occupying the redshift interval
$0.28<z<0.6$, i.e. 
\begin{equation}
{
V_{eff} = \frac{1}{N} \sum_{i=1}^{N} V_{acc,i}, 
}
\end{equation}
where the index $i$ is summed over the 466 early--type galaxies for
which $V_{acc}$ is computed.  Effective survey volumes computed for
each galaxy formation model presented in Table \ref{tab_lf_models} are
displayed in Table \ref{tab_eff_vols}.
\begin{table}
\caption{Effective spectroscopic survey volume computed for each
galaxy formation model presented in Table \ref{tab_lf_models}.} 
\label{tab_eff_vols}
\begin{tabular}{lc}
Model & $V_{eff} (\times 10^5 \, h^{-3} \, {\rm Mpc}^3)$ \\
1 & 7.1 \\
2 & 7.7 \\
3 & 6.5 \\
4 & 5.5 \\
5 & 5.3 \\
6 & 4.7 \\
\end{tabular}
\end{table}
The effective survey volume may be compared to the space density of
rich galaxy clusters in order to constrain the contribution of the
field early--type galaxy sample by cluster early--type galaxies. The
space density of rich ($R \ge 1$) Abell clusters at redshifts $z <
0.6$ is $5.6 \times 10^{-6} \, h^3 \, {\rm Mpc}^{-3}$ (Postman et
al. 1996). Comparing this space density to the effective survey
volumes for a spatially flat universe indicates that the spectroscopic
early--type galaxy sample may contain four rich galaxy clusters. At a
redshift $z=0.4$, the photometric selection criteria will sample the
cluster early-type galaxy magnitude distribution down to $\sim m_{10}$
and we therefore assume a cluster contribution rate in the full
photometric sample of $\sim 10${\%}. However, the strict lower limit
on the spatial sampling of galaxy targets imposed by the 2dF
instrument (greater than 30 arcseconds) implies that it is unlikely
that the presence of a rich cluster in the spectroscopic target fields
will contribute more than a few galaxies to the sample. This
conclusion is supported by the absence in the spectroscopic sample of
any groups of 5 to 10 galaxies within 2000 kms$^{-1}$ velocity
intervals and within a projected radius of 5 arcminutes -- thus
implying a cluster contribution rate for the spectroscopic subsample
of $<5-10${\%}.

\subsection[]{The luminosity function: the STY maximum likelihood
estimator}
\label{sec_sty}

The STY estimator (Sandage, Tammann \& Yahil 1979) is unbiased by density
variations and, although the density normalisation is not constrained,
the method provides a robust determination of the shape of the
luminosity function.  The galaxy luminosity function may be defined via
the number of galaxies with magnitudes in the interval $M$ to
$M+{\rm{d}}M$ within a volume element d{\bf x}$^3$ to be
$\phi(M,{\bmath {\rm{x}}}) \, {\rm{d}}M \, {\rm{d}}{\bmath
{\rm{x}}}^3$. Based on the results of the previous section we assume
that $\phi(M,{\bmath{\rm{x}}})=\phi(M)$, where the given functional
form of the luminosity function is characterised by a constant space
density $\phi^{\ast}$ per unit volume and magnitude interval. We
further assume that the shape parameters describing $\phi(M)$ do not
vary with redshift. Thus, one may express the probability density that
a sample galaxy observed at a redshift $z$ will display an absolute
magnitude $M$ as
\begin{equation}
{ 
p(M|z) \propto {\frac{\phi(M) \,
f(m)\,S(z)\,({\rm d}V/{\rm d}z)}
{\int_{z_{min}}^{z_{max}}\int_{-\infty}^{\infty} \phi(M^{\prime}) \,
f(m^{\prime}) \, S(z^{\prime}) \, ({\rm d}V/{\rm d}z^{\prime})\,
{\rm{d}}M^{\prime} \, {\rm d}z^{\prime}}}.  
}
\end{equation}
The factor $S(z)$ describes the colour--redshift selection probability
(Equation \ref{eqn_colour_select}) and ${\rm d}V/{\rm d}z$ is the
co--moving differential volume element per unit solid angle. The factor $f(m)$
indicates the fraction of galaxies entering the early--type galaxy
sample as a function of apparent magnitude, $m$
\begin{eqnarray}
f(m) &=& 0 \;\;\;\;\;\;\;\;\; {\mbox{where}} \: M>M_{faint}(z) \\\nonumber
&& \;\;\;\;\;\;\;\;\;\;\; {\mbox{or}} \: M<M_{bright}(z) \\\nonumber 
f(m) &=& c(m) \:\:\:\: {\mbox{where}} \:
M_{bright}(z)<M<M_{faint}(z).\nonumber  
\end{eqnarray}
The factor $c(m)$ is the redshift incompleteness as a function of
apparent magnitude $m_{i}$. The relation between $m_{i}$ and
rest--frame absolute $V$--band magnitudes is given by equations
\ref{eqn_magz} and \ref{eqn_abs}.

The luminosity function is parameterised in terms of a Gaussian
distribution of characteristic magnitude $M^{\ast}$ and width
$\sigma_M$. The photometric errors are incorporated by convolving the
intrinsic form of the luminosity function, $\phi_{int}(M)$, with a
Gaussian function of standard deviation equal to $\sigma_{obs}$, the
observed $1\sigma$ $i$--band magnitude error of each galaxy, to
generate an observed luminosity function,
\begin{equation}
{
\phi_{obs} = {\frac{1}{\sqrt{2\pi}\sigma_{obs}}}
\int_{-\infty}^{\infty} \phi_{int}(M^{\prime}) \, \exp \left (
{\frac{-{(M-M^{\prime})}^2}{2\sigma_{obs}^2}}\right )
{\rm{d}}M^{\prime}.
}
\end{equation}
The maximum likelihood procedure identifies values of $M^{\ast}$ and
$\sigma_M$ that maximise the log likelihood equation over the $k=1, N$
sample galaxies drawn from the redshift interval $0.28<z<0.6$, i.e.
\begin{equation}
{
\ln {\mathcal L} = \sum_{k=1}^{N} \ln p \, (M_i|z_i) \, .
}
\end{equation}
Confidence intervals in the $M^{\ast}$ versus $\sigma_M$ may be
displayed by computing contours of equal likelihood according to
Equation \ref{eqn_max_err}. The characteristic
density is estimated by normalising the integrated number of galaxies
predicted for a given luminosity function model to the total number of
early--type galaxies observed in the full photometric sample (Section 2,
Paper I; Section 2), i.e.
\begin{multline}
N_{total} = {\rm{d}}\Omega \int_{z_{min}}^{z_{max}}
\int_{m_{bright}}^{m_{faint}} \phi_{obs}(m(M,z)) \\ \times \, 
S(z) \, ({\rm{d}}V/{\rm{d}}z) \, {\rm{d}}m \, {\rm{d}}z.
\label{eqn_phi_calc}
\end{multline}
The apparent magnitude of an early--type galaxy of absolute
$V$--magnitude, $M$ and redshift, $z$, is calculated employing
Equations \ref{eqn_magz} and \ref{eqn_abs} assuming a mean Galactic
extinction of $E(B-V)=0.0294$, calculated using the full photometric
sample.

\subsection[]{Results and discussion}

\begin{figure}
\psfig{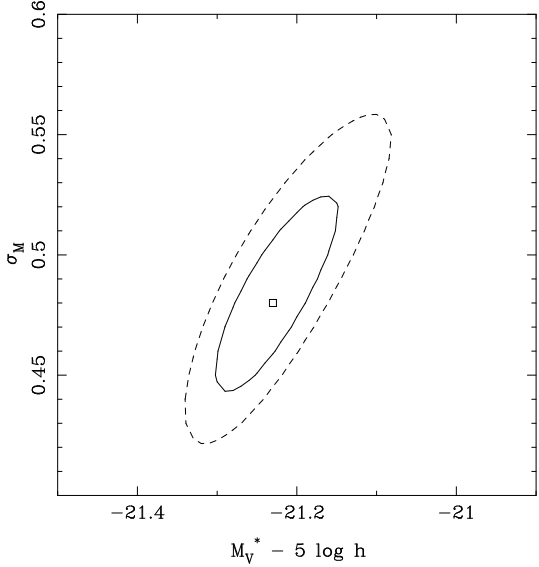}
\caption{Contours of relative likelihood corresponding to 1 (solid)
and $2\sigma$ (dashed) joint error distributions for luminosity
function parameters $M^{\ast}_V-5\log h $ and $\sigma_M$. Contours are
computed for Model 1 (Table \ref{tab_lf_results}). The open square
indicates the formal maximum.}
\label{lf_cont}
\end{figure}

The results of the maximum likelihood analysis for each spectral
evolution and cosmological model combination described in Table
\ref{tab_lf_models} are given in Table \ref{tab_lf_results}. Figure
\ref{lf_cont} displays error contours associated with the luminosity
function parameters determined for Model 1.
Error contours represent the relative uncertainty associated with a
given combination of spectral evolution and cosmological model and do
not account for uncertainties in the adopted spectral evolution model
or cosmological parameters. Errors are thus likely to be underestimated
and the luminosity function parameters returned by the maximum
likelihood procedure should be considered solely as representations of
the data over a limited range of absolute magnitude (Lilly {\etal}
1995). Given the limited range of absolute magnitude and the strong
correlation between the errors of the parameters of the Gaussian fit to
the luminosity function, the integrated luminosity density is a far more
robust quantity. Consideration of the integrated luminosity density has
the further advantage that results based on fitting different functional
forms for the luminosity function may also be intercompared. 

A significant degree of uncertainty can arise when comparing luminosity
functions of galaxies defined using different data sets or selected
according to different selection criteria. It is necessary to establish
both that the same type of objects are included in the determination of
the luminosity functions and that the transformations between measured
quantities, particularly absolute magnitudes, are well understood. The
situation is compounded, as here, when a comparison between luminosity
functions of galaxy samples at very different redshifts is attempted.
Fortunately, the recent availability of the Sloan Digital Sky Survey
(SDSS) Early Data Release (EDR) (Stoughton et al. 2002) and the
determination of the luminosity function for early--type galaxies
(Eisenstein et al.  2001, Bernardi et al. 2002), allow a relatively
direct comparison between our results at $z\simeq 0.4$ and those
pertaining to the present day to be made. Bernardi et al. present
luminosity functions in the SDSS $g^*r^*i^*z^*$ bands for a
sample of early--type galaxies with $\langle z \rangle
\simeq 0.1$.  Early--type galaxies within the SDSS are defined
according to a combination of morphological (concentration and surface
brightness) and spectral (no detectable emission lines and projected
velocity dispersion greater than the effective instrumental limit)
criteria. The resulting galaxy sample possesses a well--defined mean
colour--redshift locus consistent with a uniformly old stellar
population, combined with a small Gaussian dispersion about the mean
relation. Therefore, in the absence of additional star--formation
events, the $z \simeq 0.4$ early--type galaxy population will evolve
passively to form early--type galaxies that satisfy the SDSS selection
criteria.
\begin{table}
\caption{Gaussian luminosity function parameters estimated via STY
maximum likelihood procedure. Parameters are valid over the
approximate absolute magnitude interval $-22.5<M_V-5\log h<-20.5$. The
$2\sigma$ single parameter errors are $-{M_V^{\ast}}^{+0.11}_{-0.15}$
for characteristic magnitude and ${\sigma_M}^{+0.08}_{-0.06}$ for the
width term. The characteristic space density, $\phi^{\ast}$, is
computed via Equation \ref{eqn_phi_calc} for the central values of
$M_V^{\ast}-5\log h$ and $\sigma_M$ determined for each model and is
expressed units of $10^{-3} \, h^3 \, {\rm{Mpc}}^{-3}
{\rm{mag}}^{-1}$. The $2\sigma$ uncertainty in the characteristic
space density expresses the variation in $\phi^{\ast}$ associated with
the corresponding $2\sigma$ variation in the luminosity function shape
parameters. This source of uncertainty dominates over the Poisson
uncertainty associated with the photometric sample size. 
}
\label{tab_lf_results}
\begin{tabular}{cccc}
Model & $-M_V^{\ast}+5\log h$ & $\sigma_M$ & $\phi^{\ast}$ \\
1 & 21.28 & 0.47 & $1.50 \pm 0.25$ \\
2 & 21.23 & 0.48 & $1.46 \pm 0.25$ \\
3 & 21.28 & 0.50 & $1.59 \pm 0.23$ \\   
4 & 21.12 & 0.47 & $1.86 \pm 0.29$  \\ 
5 & 21.10 & 0.46 & $1.96 \pm 0.32$  \\ 
6 & 21.10 & 0.49 & $2.17 \pm 0.32$  \\ 
\end{tabular}
\end{table}

%
%

The Gaussian luminosity function parameters of Bernardi et al. in the
$g^{\ast}$--band are transformed to rest--frame $V$--band values at a
redshift $z=0.4$. The redshift $z=0$ $g^{\ast}$--band characteristic
magnitude is corrected for passive luminosity evolution to a redshift
$z=0.4$ employing the $Q$--parameter estimated by Bernardi et al. The
zero--point transformation between the $g^*$--band absolute magnitudes
on the SDSS AB system of Bernardi et al. to those in the $V$--band on
the Vega system is readily accomplished using the appropriate
transmission functions and the early--type galaxy template of Coleman,
Wu \& Weedman (1980) employed by Bernardi et al. Application of
transformations to account for aperture to total magnitudes and a
residual zero--point difference result in a total shift of UKST
$i$--band magnitudes of $-0.48\,$mag relative to the SDSS system (see
Appendix \ref{app_sdss_comp} for a comprehensive discussion). The two
samples are then compared in terms of the $V$--band luminosity density
integrated over the common absolute magnitude interval sampled by each
survey,  $-23 \la M_V^{\ast} - 5 \log h \la -21$, i.e.
\begin{equation}
{
\rho_L = \int_{M_{bright}}^{M_{faint}} \phi(M) \,
10^{-0.4(M-M_{{\odot},V})} \, {\rm d}M,
}
\label{eqn_lum_den}
\end{equation}
which is expressed in units of $\times 10^7 \, h^2 \, L_{{\odot},V} \,
{\rm Mpc}^{-3}$, assuming that the $V$--band absolute magnitude of the
Sun is $M_{{\odot},V}=4.83$. 
\begin{table}
\caption{Luminosity density values calculated for SDSS and $z=0.4$
early--type galaxy samples according to Equation
\ref{eqn_lum_den}. Values are expressed in units of $\times 10^7 \,
h^2 \, L_{{\odot},V} \, {\rm Mpc}^{-3}$ (see text). The Poisson error
for each values is approximately 1{\%}. Note that the SDSS luminosity
function values from which the luminosity density is generated are
presented by Bernardi et al. for the cosmological model
$\Omega_M=0.3$, $\Omega_{\Lambda}=0.7$ and $H_0=70$ km s$^{-1}$
Mpc$^{-1}$. Hence, for consistency, the SDSS figure is compared only
to Models 1--3 described in Table \ref{tab_lf_models}.}
\label{tab_ld}
\begin{tabular}{ccc}
Model & $\rho_L$ & fraction compared to SDSS \\
SDSS & 7.21 & -- \\
1 & 6.87 & 0.95 \\
2 & 6.38 & 0.88 \\
3 & 7.19 & 1.00 \\
\end{tabular}
\end{table}
Comparison of the luminosity density determined for the SDSS and
$z=0.4$ samples (Table \ref{tab_ld}) indicates that, within the
uncertainties imposed by the spectro--photometric model employed to
describe the $z=0.4$ early--type galaxy sample and the transformations
applied to place the two samples on a common photometric system, there
is no significant shortfall in luminous field early--type galaxies at
redshift $z=0.4$ compared to $z=0.1$.

\section{Discussion}

The properties of luminous, field early--type galaxies at redshifts
$z<1$ in principle provide an important test of the hierarchical model
of galaxy formation. We have presented an analysis of the mean star
formation history and space density of a sample of luminous field
early--type galaxies selected over the redshift interval $0.3 \la z \la
0.6$.

Overall, the mean star formation history of the sample is characterised
by an apparently old ($z_f>1$), solar to slightly above--solar
metallicity luminosity--weighted stellar population that has evolved
passively since the formation epoch. The colour--redshift evolution,
mean absorption line properties and mean spectrum of the sample present
a consistent picture of an old, quiescent stellar population. The exact
range of formation redshift and metallicity permitted depends upon the
assumed time dependence of the major star formation event and the
cosmological model via the look--back time versus redshift relation.
The mean properties of the sample are markedly similar to the
properties of morphologically--selected luminous elliptical galaxies in
rich cluster environments at redshifts $z<1$ (Ferreras {\etal} 1999).
Though neither result in isolation constrains the extent to which
early--type galaxies in field or cluster environments represent a
co--eval or co--metal population, the broad similarity between the star
formation history of the dominant stellar mass component in such
galaxies is consistent with similar formation conditions for each
population.

Approximately one--quarter of the early--type galaxy sample displays
detectable [OII]3727 emission consistent with on--going star formation
rates $\la 1.5 h^2$ M$_{\odot}$ yr$^{-1}$. Consideration of the number
redshift distribution of [OII] 3727 emitting galaxies and the effects
of a recent burst of star formation upon the colour selection of such
galaxies, limits the typical stellar mass content of such star
formation events to $<1${\%} of the galaxy mass for a burst occurring
at redshift $z=1$. However, the space density of the strongest [OII]
3727 emitting galaxies in the sample displays a marked increase at
redshifts $z \ga 0.4$. Although further detailed observations are
required to determine the true nature of such star formation events,
the star formation rate associated with each event combined with the
increase in space density at higher redshift is consistent with an
increasing accretion rate of low mass, gas rich galaxies with
increasing look--back time.

Comparison of the luminosity density of the $z = 0.4$ early--type
galaxy sample to the sample of $z = 0.1$ early--type galaxies from the
SDSS shows no evidence of significant change in the luminosity density
of luminous field early--type galaxies between these two epochs. In
common with results obtained for morphologically selected samples of
elliptical galaxies obtained over similar redshift intervals (Schade
{\etal} 1999), these results show no evidence that significant numbers
of luminous early--type/elliptical galaxies have formed via merging at
redshifts $z<1$. The hypothesis that luminous field early--type
galaxies formed via the hierarchical merging of gas rich disk galaxies
is not rejected by our results. However, it does seem clear that the
bulk of star formation associated with the formation of such galaxies
occurred at redshifts $z>1$ and the assembly of the most massive
early--type galaxies in the field, via any postulated merging process,
was largely complete by redshift $z\simeq 0.5$.

\section*{Acknowledgments}
JPW acknowledges the support of a PPARC research studentship and
financial support from IoA, Cambridge.  The project would not have
been possible without the data and analysis facilities provided by the
Starlink Project which is run by CCLRC on behalf of PPARC. We further
thank an anonymous referee for providing a careful and constructive
report.

\appendix

\section[]{Modeling the distribution of $b_{\rm J}ori$ colours}
\label{app_mod_col}

For galaxies at a redshift $z$, the distribution of $b_{\rm J}-or$ versus 
$or-i$ colours may be described by a two--dimensional Gaussian
function, $G$, of the form,

\begin{multline}
G(u,v \, | \, z) = {\frac{1}{2 \pi \, \sigma_u \, \sigma_v}} \exp \left[
{\frac{-(u-u_m)^2}{2 \sigma_u^2}} \right] \\ \times \exp \left[
{\frac{-(v-v_m)^2}{2 \sigma_v^2}} \right]. 
\end{multline}

{\noindent}Where

\begin{eqnarray}
u &=& (or-i) - (b_{\rm J}-or) \\
u_m &=& {(or-i)}_{model}(z) - {(b_{\rm J}-or)}_{model}(z) \\ \nonumber
\sigma_u &=& (2\sigma_{or}^2 + \sigma_{b_{\rm J}}^2 +
\sigma_{i}^2)^{1/2} \nonumber
\end{eqnarray}

{\noindent}and

\begin{eqnarray}
v &=& (or-i) + (b_{\rm J}-or) \\
v_m &=& {(or-i)}_{model}(z) + {(b_{\rm J}-or)}_{model}(z) \\ \nonumber
\sigma_v &=& (\sigma_{b_{\rm J}}^2 + \sigma_{i}^2)^{1/2}\, . \nonumber
\end{eqnarray}

\begin{figure}
\psfig{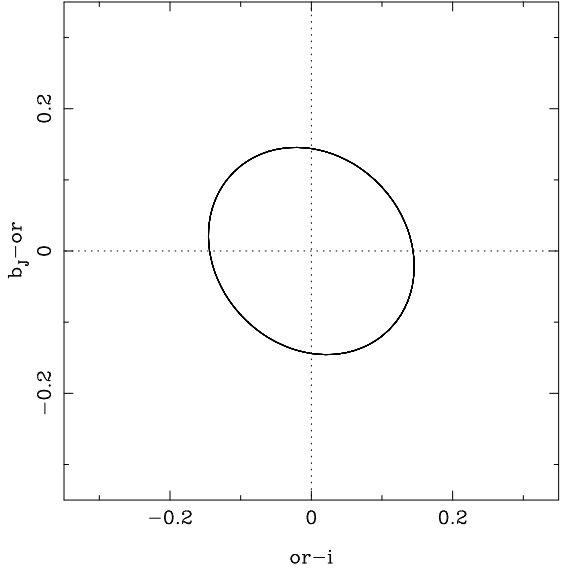}
\caption{Contour describing the $1\sigma$ probability distribution of
$b_{\rm J}-or$ versus $or-i$ colours generated by the colour distribution
function, $G$, described by the early--type galaxy sample median
photometric errors in each passband, i.e. $\sigma_{b_{\rm J}}=0.14$,
$\sigma_{or}=0.08$ and $\sigma_i=0.13$ magnitudes respectively.}
\label{example_col_dist}
\end{figure}

{\noindent}Variables employing the subscript ``model'' describe the
colour of a model early--type galaxy simulated at a redshift $z$
according to the specified spectro--photometric model. Photometric
errors in $b_{\rm J}ori$ passbands are denoted by $\sigma_{b_{\rm J}}$,
$\sigma_{or}$ and $\sigma_{i}$ respectively and are assumed to
be Gaussian in form (see Paper I; Section 2.1). Figure
\ref{example_col_dist} displays the form of the observed colour
density function $G$.

\section[]{The photometric selection function}

The photometric selection function describes the probability that a
galaxy of given $b_{\rm J}ori$ magnitude and spectral properties
described by a suitable spectral evolution model enters the sample
selection criteria as a function of redshift. The selection
probability, $S(z)$, is the
probability that a galaxy at a redshift $z$, drawn from a simulated
galaxy population characterised by the colour--redshift locus computed
from the specified spectral evolution and cosmological models together
with the photometric errors, will satisfy the sample colour selection
criteria (Table \ref{tab_photo_select}). Given the distribution of
galaxy colours drawn from a simulated population (Appendix A),
computation of the photometric selection probability is
straightforward, i.e.
\begin{multline}
S(z) = \int_{2.15}^{3.00} \int_{lim(b_{\rm J}-or)}^{1.05} G(b_{\rm J}-or,or-i \, |
\, z) \\ \times \, {\rm d}(b_{\rm J}-or) \,{\rm d}(or-i).
\label{eqn_colour_select}
\end{multline}
Where the function $lim(b_{\rm J}-or)$ describes the $or-i$ lower
integration limit as a function of $b_{\rm J}-or$ colour,
\begin{equation}
{
lim(b_{\rm J}-or) = 3.00 - (b_{\rm J}-or).
}
\end{equation}
The magnitude selection propability as a function of redshift, $P(z)$,
describes the probability that a galaxy characterised by an apparent
$i$--band magnitude versus redshift relation, $m_i(z)$ (Equation
\ref{eqn_magz}), and a $1\sigma$ photometric uncertainty, $e_i$,
satisfies the magnitude selection threshold (Table \ref{tab_photo_select})
at any given redshift, i.e.
\begin{equation}
{
P(z) = {\frac{1}{\sqrt{2\pi}e_i}} \int_{16.5}^{18.95} \exp \left (
{\frac{-{(m_i-m_i(z))}^2}{2e_i^2}}  \right )\, {\rm{d}}m_i.
}
\end{equation}
The combined photometric selection function may be expressed as the
convolution of the two probability terms, i.e.
\begin{equation}
{
W(z) = S(z) \, P(z).
}
\label{eqn_wz}
\end{equation}
Equations \ref{eqn_colour_select}--\ref{eqn_wz} describe the case
$E(B-V)=0$. For the case $E(B-V)>0$, early--type galaxy $b_{\rm J}ori$
photometry and photometric selection limits are adjusted accordingly.

\section[]{A comparison of the photometric properties of the SDSS and
$z=0.4$ early--type galaxy samples}
\label{app_sdss_comp}

Any firm conclusions regarding the evolution of the luminosity density
of the early--type galaxy population rely critically on understanding
the relationship between our sample of galaxies and that of Bernardi
et al. (2001).  The difficulties inherent in achieving such an
understanding are illustrated in the discussion of Eisenstein et
al. (2001), particularly \S4.2 and Appendix B, in linking the Luminous
Red Galaxy (LRG) sample at redshifts $z \ge 0.3$ to the SDSS MAIN
galaxy sample at lower redshifts.  Eisenstein et al. have defined
selection criteria designed to achieve the identification of a sample
of galaxies complete to a fixed absolute magnitude over an extended
range of redshift.  However, for practical reasons they are
constrained to apparently bright objects, with the result that their
absolute magnitude limit corresponds to a point in the luminosity
function where the number of galaxies is rising extremely rapidly. Our
sample, defined using straightforward apparent magnitude and colour
cuts, does not possess the desirable property of probing an absolute
magnitude range independent of redshift (Paper I; Figure 19). On the
other hand the sample extends sufficiently faint that, for redshifts $z
\la 0.5$, the absolute magnitude range extends into a much flatter
portion of the luminosity function. Specifically, our magnitude limit
corresponds on the SDSS system to $r^*_{Petro}\simeq 19.9$ compared to
the (main) Cut I of $r^*_{Petro}\simeq 19.2$ and Cut II of
$r^*_{Petro}\simeq 19.5$ employed by Eisenstein et al.. Thus, our
sample, while not probing to fixed absolute magnitude independent of
redshift, includes a much larger fraction of the early--type galaxy
population and does not involve a selection cut falling in a steeply
rising portion of the luminosity function.

The SDSS EDR contains objects in $\sim 12\,$deg$^2$ areas in two of the
seven United Kingdom Schmidt Telescope (UKST) survey fields that make
up our full photometric sample (Paper I). The overlap between the two
samples allows a direct comparison of the properties of $\sim 1000$
objects in common and also provides an empirical determination of the
transformation between the SDSS photometric system and the natural UKST
system employed throughout this paper.

The range of $g^*-r^*$ and $r^*-i^*$ colours for the 1014 objects in
common in the F855 and F864 fields are statistically indistinguishable
from the subset of galaxies that are actually included in the
Eisenstein et al. LRG. There is a similarly encouraging agreement in
the selection--colour ($c_{par}$) {\it versus} $r^*_{Petro}$
parameter--space (Eisenstein et al.; Figure 3), with an essentially
one--to--one correspondence between objects satisfying the LRG
selection criteria and those in our sample for magnitudes $r^*_{Petro}
\la 18.7$. Our simpler colour--selection criteria corresponds to a
closer to horizontal boundary in $c_{par}$ for magnitudes $18.7 \la
r^*_{Petro} \la 19.2$ and additional (intrinsically fainter at given
redshift) galaxies in our sample lie approximately $0.1\,$mag bluer in
$c_{par}$ in the faintest half--magnitude of the Eisenstein et al. Cut
I selection.

The morphological properties in the SDSS of the 1014 objects in common
are very encouraging. Less than $<3\%$ of the objects are classified
as stellar or indeterminant and $85\%$ of the sample are ``best--fit''
by a de Vaucouleurs profile. 

The magnitude zero--points used for the sample of early--type galaxies
described in Paper I were determined using CCD observations of
galaxies measured within $8\,$arcsec diameter circular
apertures. However, SDSS early--type galaxy magnitudes are presented
as ``total'' magnitudes obtained by integration of a de Vaucouleurs
surface brightness profile determined for each galaxy. Therefore, a
correction must be applied to transform the $8\,$arcsec early--type
galaxy magnitudes presented in this paper to ``total'' magnitudes. An
empirical determination of the difference between the SDSS EDR model
``total'' and $8\,$arcsec diameter aperture magnitudes for the $\simeq
1000$ galaxies in common gives a median difference of $0.32\,$mag in
the SDSS $i^*$, i.e., the ``total'' magnitudes are $0.32\,$mag
brighter than the aperture magnitudes and the magnitudes in Paper I
should be corrected brightward by this amount. Both SDSS and $z=0.4$
early--type galaxy magnitudes have been corrected for Galactic
extinction using the maps of Schlegel et al. (1998).

Finally, following the transformation of the two samples onto a common
magnitude system, i.e., either Vega--based or ABmag--based, it is
possible to compare the magnitude scales of the galaxies in the two
samples.  Specifically, the information available in the SDSS EDR for
the galaxies in common can be used to derive $8\,$arcsec diameter
aperture magnitudes to be compared to our magnitudes, also based on
$8\,$arcsec diameter measurements. The comparison for the SDSS $r^*$
and our $or$ values is excellent. The median difference between the
magnitudes (transformed to a consistent system) is $+0.02\,$mag in the
sense that our $or$ magnitudes are fainter than the SDSS values. The
offsets for the two independent fields, F855 and F864, are also in good
agreement. Considering the same comparison for the SDSS $i^*$ and our
$i$ magnitudes also produces excellent agreement between the two
independent fields but the median magnitude difference is $+0.16\,$mag,
again in the sense that our magnitudes are fainter than the SDSS
values. This, well--determined, offset in the magnitudes is
uncomfortably large. Eisenstein et al. discuss the question of residual
uncertainties in the zero--points of the SDSS photometric bands and
conclude that there could be errors in the preliminary calibrations of
the SDSS bands, but it seems very unlikely that such errors would
exceed $0.05\,$mag in the SDSS $i$--band.  Our own zero--pointing
procedure (Paper I; \S 2.2) is uncertain to $\simeq 0.05\,$mag and a
systematic error of $\simeq 0.1\,$mag is not out of the question.
However, the difference of $+0.16\,$mag appears large given the quoted
uncertainties and we have no entirely satisfactory explanation for the
difference. The difference in the $i$--band magnitude scale is
particularly important because our absolute magnitudes are calculated
using the $i$--band apparent magnitudes. Ideally, it is desirable to
undertake absolute determinations of luminosity functions at different
redshifts, but in practice the question of whether there has been any
significant evolution of the luminosity function with redshift relies
on a differential measure. Thus, when comparing our determination of
the luminosity density of early--type galaxies at $z\simeq 0.4$ to that
at lower redshifts from Bernardi et al. we will use a value calculated
using $i$--band magnitudes shifted brightward by $0.16\,$mag.

To place the SDSS and $z=0.4$ early--type galaxy samples on a common
photometric scale, we therefore apply a global shift of $-0.48\,$mag
to all early--type galaxy UKST $i$--band magnitudes.

\bsp

\label{lastpage}

\end{document}